\documentclass[aps,prd,endfloats*,showpacs]{revtex4}
\usepackage{graphicx}
\newcommand{\be}{\begin{equation}}
\newcommand{\ee}{\end{equation}}
\newcommand{\ba}{\begin{eqnarray}}
\newcommand{\ea}{\end{eqnarray}}
\newcommand{\sighat}{\hat{\sigma}}
\newcommand{\shat}{\hat{s}}
\newcommand{\that}{\hat{t}}
\newcommand{\uhat}{\hat{u}}
\newcommand{\qbar}{\overline q}
\newcommand{\np}{{Nuclear Phys.}}

\begin{document}
\title{Parameterizations of Invariant Meson Production Cross Sections}
\author{Alfred Tang}
\email{atang@alum.mit.edu}
\affiliation{Department of Physics, Baylor University,
P. O. Box 97316, Waco, TX 76798-7316}
\author{John W. Norbury}
\email{norbury@uwm.edu}
\affiliation{Physics Department, University of Wisconsin-Milwaukee,
P. O. Box 413, Milwaukee, WI 53201.}
\date{\today}

\begin{abstract}
The Lund string fragmentation model is applied in a non-perturbative
calculation of
the invariant production cross sections of pions from proton-proton
collisions in the soft $p_T$ region.
Invariant production cross sections of pions and kaons from proton-proton
collisions in the hard $p_T$ region
are calculated from the Feynman-Field perturbative QCD parton model.
Parameterizations of these invariant production cross sections are
described.
\end{abstract}
\maketitle

\section{Introduction}
This work was originally motivated by the need of parameterized meson
production cross sections of $pp\to hX$ reactions ($p$ for primary proton, $h$
for identified hadron production and $X$ for unidentified hadron production)
for a NASA nuclear
transport code called \texttt{HZETRN}~\cite{wilson95}.  The parameterization
scheme presented here parameterizes theory.  Experimental data are used mostly
as checks.  The parameterizations are based on
two main calculations---a non-perturbative QCD string fragmentation Lund model
calculation in the soft $p_T$ region $(p_T<1\,\rm GeV)$ and a perturbative
QCD Feynman-Field parton model calculation in the hard $p_T$ region
$(p_T\ge1\,\rm GeV)$.  The threshold of $1\,\rm GeV$ separating the soft
and hard $p_T$ regions is chosen to be the proton mass $m_p=938\,\rm MeV$.
Descriptions of the Lund and Feynman-Field models can be found in
original sources~\cite{andersson98,field} and a review paper~\cite{appen}.
Both the Lund model and the Feynman-Field model are stochastic models
and are not explicitly quantized.  In this sense, both are phenomenological
models.  The goal of this work is to calculate $E\,d^3\sigma/dp^3$
with sufficient accuracy using reasonable theoretical models so that reliable
parameterizations of cross section formulas can be determined.
The Feynman-Field model is implemented numerically by the Monte Carlo
integration package \texttt{VEGAS}~\cite{nr97,lepage80}.  The
invariant cross sections are parameterized for pions in the soft
$p_T$ region and for pions and kaons in the hard $p_T$ regions.

\section{Pion Production in the Soft $p_T$ Region}
String fragmentation models such as the Lund Model fit experimental data
well.  String theory reproduces the linear potential predicted
by non-perturbative QCD as in lattice gauge field theory.  These
observations hint at the possibility that QCD string may be
conducive to solving non-perturbative QCD.  Although
the Lund model is a $(1+1)$ model, it reproduces the essential dynamics of
the system as long as information on angular momentum is not needed.
Typically the Lund model is implemented numerically using Monte Carlo
simulation in \texttt{JETSET} and \texttt{PYTHIA}~\cite{sjostrand01}.
This section shows how to calculate meson production cross section formulas
analytically in the non-perturbative QCD region using the Lund
model.  Special attention is given to the invariant production cross
sections of pions from proton-proton scattering in the soft $p_T$ region
$(p_T<1\,\rm GeV)$ where
non-perturbative effects dominate.  The results of the string model are
compared against inclusive pion production cross section data of
proton-proton collision.

\begin{figure}
\includegraphics[scale=0.8]{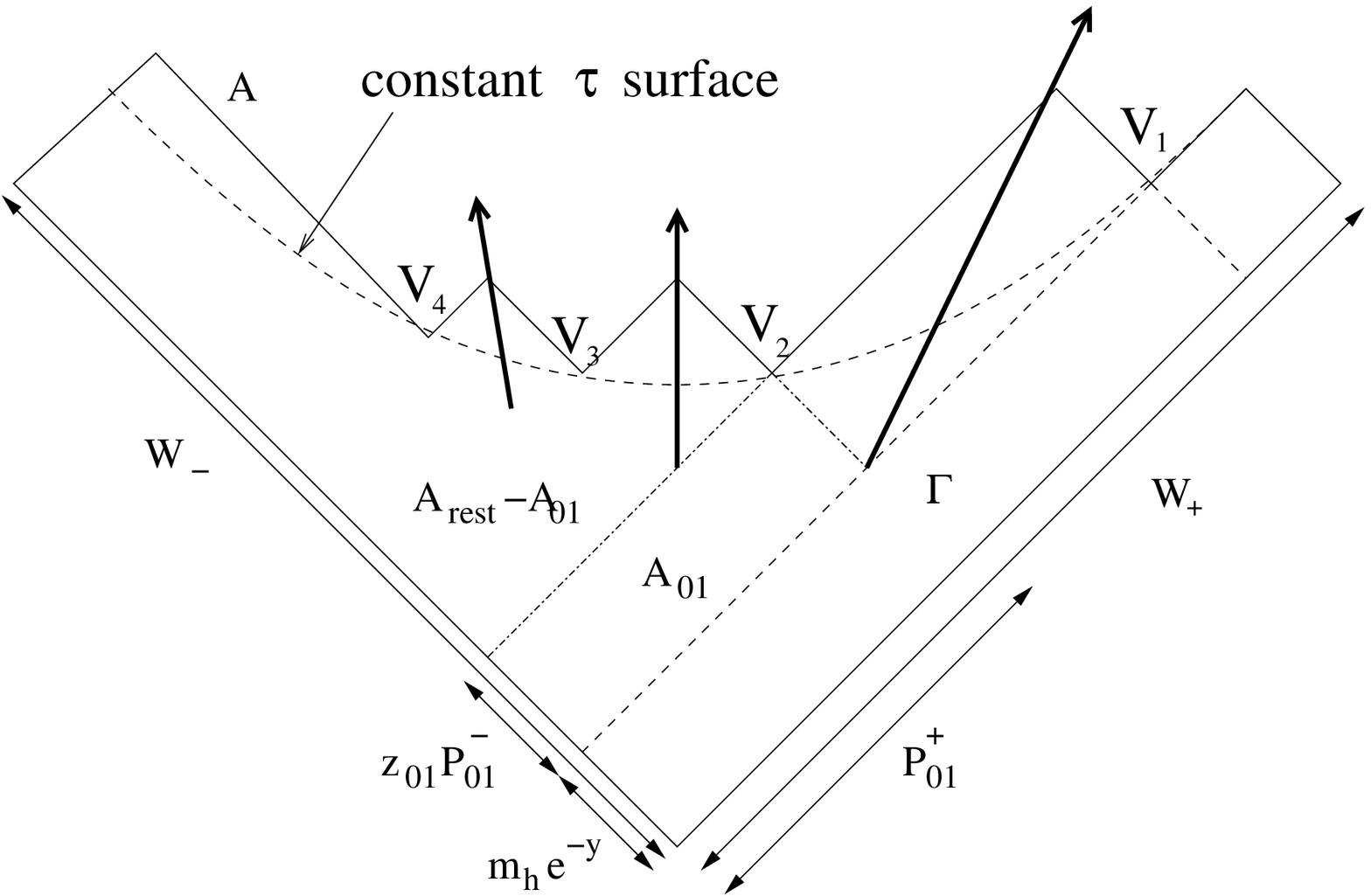}
\caption{\label{pic}
A spacetime diagram of the $(1+1)$ Lund string fragmentation process.
The vertices $V_i$ denote the spacetime points of the creation of virtual
quark pairs.  The quarks are massless so that their trajectories are
light-like.  The arrows indicate the directions of the trajectories of the
produced hadrons.}
\end{figure}

This section uses the same notations of Reference~\cite{andersson98}.
The basic result of the Lund Model is the ``area law'' which is
summarized as~\cite{andersson98}
\be
dP=ds\,{dz\over z}\,(1-z)^a\,
\prod^n_{j=1}N\,dp^2_{0j}\,\delta^+(p_{0j}^2-m^2)\,
\delta\left(p_{rest}-\sum^n_{j=1}p_{0j}\right)\,e^{-bA}.\label{ar}
\ee
$P$ is the probability of the string fragmentation process,
$b$ and $N$ are constants and $s$ is the total energy square of $n$ produced
mesons.  Lightcone coordinates are used
in the Lund model, $x_\pm=t\pm x$ and $p_\pm=E\pm p$.  The
momentum $p_{rest}=(W_-,W_+)$ is that of the parent virtual quark pair
and $p_{0j}$ is that of the $j$-th rank meson.  Momentum transfer
is $-t=-q^2=W_-W_+$ where $t$ is a Mandelstam variable.  The mass of the
system is $m$.  The area
$A=\Gamma+A_{rest}$ of the polygon in Fig.~\ref{pic} has a geometric
interpretation as the residual energy of the
virtual quark pair following the fragmentation process.
$\Gamma=\kappa^2 x_+x_-$, $A$ and $A_{rest}$ are Lorentz invariant kinematic
variables.  The symbol $z=\sum_j z_{oj}$ denotes the sum of the fractions
$z_{0j}$ of lightcone energies
of the produced mesons along the $x_+$ direction.  The quantity $\Gamma$
defines the surface of constant proper time along which the string is broken.
The area law captured in Eq.~(\ref{ar}) is usually incorporated into Monte
Carlo simulation programs such as \texttt{JETSET} and
\texttt{PYTHIA}~\cite{sjostrand01}.
Monte Carlo subroutines are too slow for nuclear transport codes.
Computational constraints
motivate the search for parameterized production cross section formulas.

\begin{figure}
\includegraphics[scale=0.6]{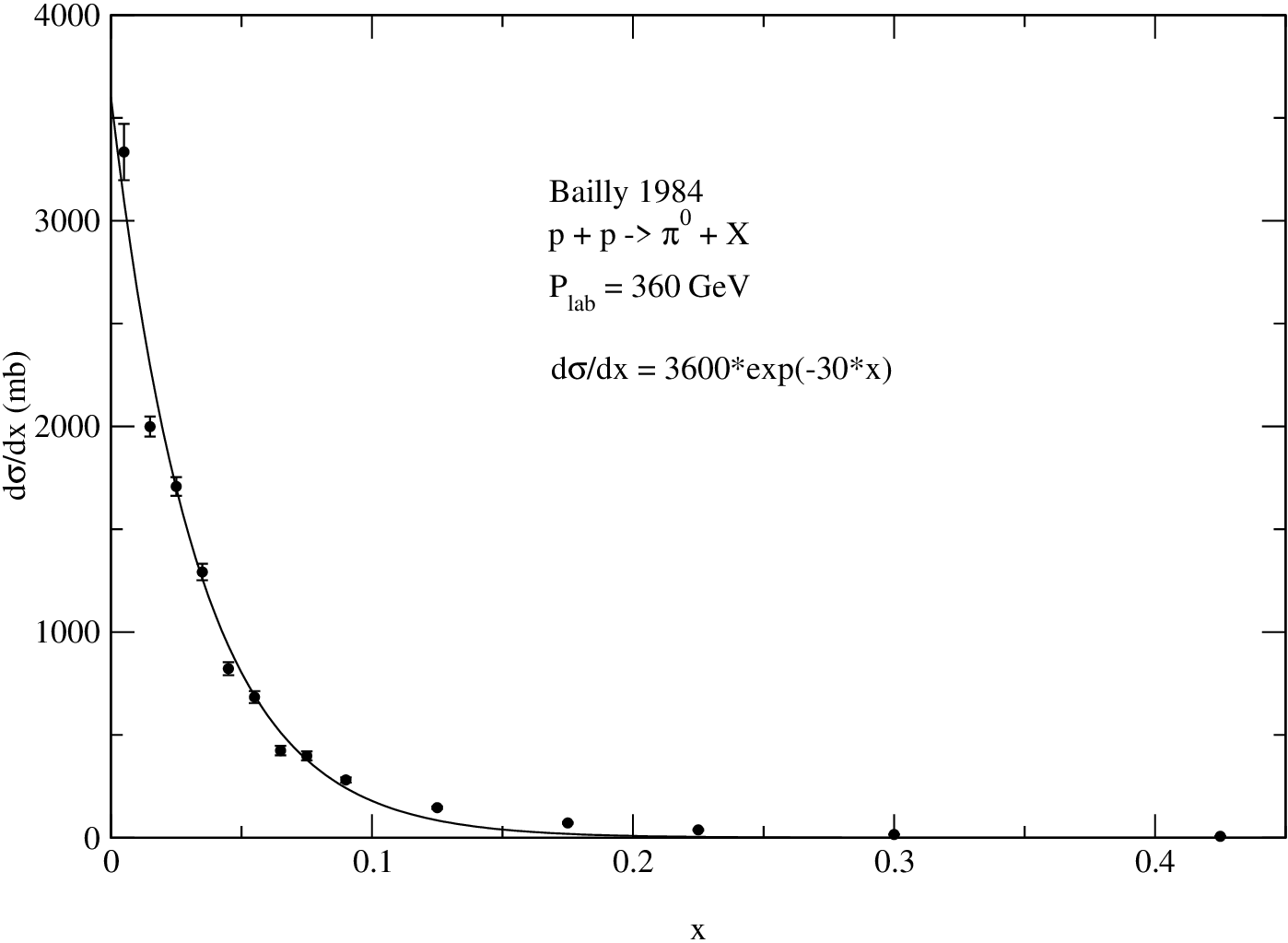}
\caption{\label{ba84}
Comparison between experimental $d\sigma/dx$ and the string model result.
The constants used in the exponential function are chosen to fit the data
and are not parameterizations.  The data are published in
Reference~\cite{bailly84}.}
\end{figure}

\begin{figure}
\includegraphics[scale=0.6]{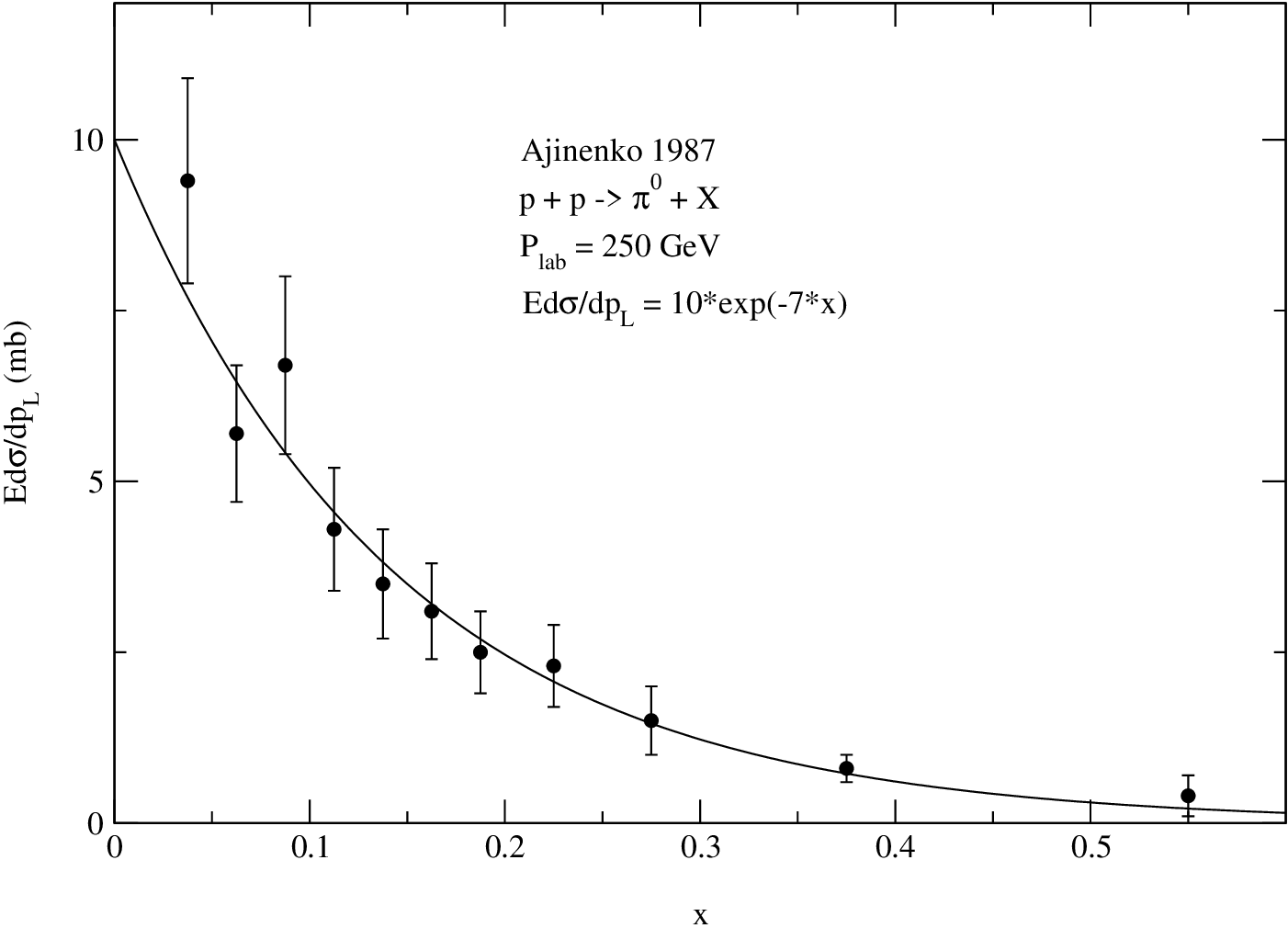}
\caption{\label{aj87}
Comparison between experimental $E\,d\sigma/dp_L$ and the string model result.
The constants used in the exponential function are chosen to fit the data
and are not parameterizations.  The data are published in
Reference~\cite{ajinenko87}.}
\end{figure}

By taking $dP=d\sigma/\sigma_0$, where $\sigma_0$ is the total cross section,
and using the identities found in References~\cite{andersson98,appen},
Eq.~(\ref{ar}) can be rewiritten as
\be
d\sigma=\sigma_0\,ds\,dz\,(1-z)^a\,e^{-b\Gamma}\,
\delta\left(z-\sum^n_{j=1}z_{oj}\right)\,
\delta\left(s-\sum^n_{j=1}{m^2z\over z_{oj}}\right)\,
\prod^n_{j=1}N\,{dz_{oj}\over z_{oj}}\,e^{-bA_{rest}},\label{ar2}
\ee
where $z_{oj}$ is the fraction of
the parent quark momentum $p_+$ that goes
into the momentum $p_{oj+}$ of $j$-th rank meson.  For $n>1$, the area 
$A_{rest}$ can be divided into $n$ rectangles so that
\be
A_{rest}=\sum^n_{j=1}A_{oj}.
\ee
Fig.~\ref{pic} illustrates the first of such $n$ rectangles and the
geometric properties of the measures of its sides.  By utilizing the
relation
\be
p_{oj+}={\Gamma\over p_{oj-}},
\ee
it can be shown that
\be
A_{oj}=z_{oj}\,\Gamma.
\ee
When $n=1$, there are 2 vertices and $A_{rest}=z_{o1}\Gamma$.  In this case,
Eq.~(\ref{ar2}) can be integrated over $s$ and $z$ to give
\be
d\sigma=\sigma_0\,N\,(1-z)^a\,e^{-b(1+z)\Gamma}\,{dz\over z}.
\ee
It is easy to show that the Feynman variable $x=p_z/(p_z)_{max}$ is equal to
$z$ when there is only 1 produced meson so that $dx=dz$.  The
differential cross section is found to be
\be
{d\sigma\over dx}=\sigma_0\,N\,{(1-z)^a\over z}\,e^{-b(1+z)\Gamma}.
\label{ar3}
\ee
The invariant quantity
\be
\Gamma=m_h\,e^{\mp y}\left(W_\pm-m_h\,e^{\pm y}\right),\label{gamma}
\ee
where $m_h$ is the mass the produced meson,
can be understood intuitively through geometrical considerations by focusing
on the rectangle in the spacetime diagram next to the first rank vertex $V_1$
in Fig.~\ref{pic}.  Generally, the lightcone energies of the parent quarks
can be taken to be the momentum transfer $W_-=W_+=q$.
The feynman variable can be defined as
\be
x\equiv{p_z\over (p_z)_{max}}={2m_h\sinh y\over q}.\label{feyn2}
\ee
With Eqs.~(\ref{gamma}) and (\ref{feyn2}), Eq.~(\ref{ar3}) can be rewritten as
\ba
{d\sigma\over dx}&=&\left[\sigma_0\,N\,{(1-z)^a\over z}
\,e^{-b(1+z)m_h\,e^{-y}(q-m_h\,e^{-y})}\right]\,e^{-b(1+z)q(q-m_h\,e^{-y})x}\\
&\simeq&\left[\sigma_0\,N\,{(1-z)^a\over z}\,e^{-b(1+z)m_h(q-m_h)}\right]\,
e^{-b(1+z)q(q-m_h)x}\label{expo_a}\\
&\equiv& C\,e^{-Dx}.
\label{expo}
\ea
Eq.~(\ref{expo_a}) is obtained by setting $y=0$.  In general, experimental
data average over rapidity that typically centers around zero such that
$y=0$ is a reasonable simplification.  In addition, average values of
internal variables $z$ and $q$ are used in Eq.~(\ref{expo_a}) so that
$C$ and $D$ can be treated as
constants when comparing with experimental data.
The form of Eq.~(\ref{expo}) is consistent with
experimental data as shown in Figs.~\ref{ba84} and \ref{aj87}.

When $n=2$, Eq.~(\ref{ar2}) is transformed as follow by integrating over
$z_{o1}$ and $z_{o2}$ and letting $\xi\equiv z_{o1}(z-z_{o1})$:
\ba
d\sigma&=&\sigma_0\,N\,(1-z)^a\,dz\,\delta(z-z_{o1}-z_{o2})\,
ds\,\delta\left(s-{m^2z\over z_{o1}}-{m^2z\over z_{o2}}\right)\nonumber\\
&&\quad\times
{dz_{o1}\over z_{o1}}\,{dz_{o2}\over z_{o2}}\,e^{-b\Gamma}\,
e^{-b(z_{01}+z_{o2})\Gamma}\nonumber\\
&=&\sigma_0\,N\,(1-z)^a\,dz\,
ds\,\delta\left(s-{m^2z\over z_{o1}}-{m^2z\over z-z_{o1}}\right)\,
{dz_{o1}\over z_{o1}(z-z_{o1})}\,e^{-b(1+z)\Gamma}\nonumber\\
&=&\sigma_0\,N\,(1-z)^a\,dz\,
ds\,\delta\left(s-{m^2z\over\xi}\right)\,
{d\xi\over\xi\sqrt{z^2-4\xi}}\,e^{-b(1+z)\Gamma}\nonumber\\
&=&\sigma_0\,N\,(1-z)^a\,dz\,ds\,
{1\over s\sqrt{z^2-{4m^2\over s}}}\,\,e^{-b(1+z)\Gamma}.\label{ar4}
\ea
One of the typical assumptions of the form of momentum transfer in QCD is
$-t=-q^2=4p_T^2$.  With the relations
$s\simeq -t-(1+z)\Gamma=4p_T^2-(1+z)\Gamma$,
we obtain
\be
ds=4dp_T^2,\label{ds}
\ee
by keeping $z$ and $\Gamma$ constant.  On the lightcone, $p_z=E$ so that
\be
dz={dE\over E}={dp_z\over E}.\label{dz}
\ee
Eqs.~(\ref{ds}) and (\ref{dz}) together give the relation
\be
dz\,ds=4\,{dp^2_T\,dp_z\over E}=4\,{dp^3\over E}.\label{dp3E}
\ee
By combining Eqs.~(\ref{ar4}) and (\ref{dp3E}), the invariant production
cross section can finally be shown to be
\be
E\,{d^3\sigma\over dp^3}=
{4N\,(1-z)^a\,\sigma_0\over\sqrt{s^2z^2-4s\,m^2}}\,e^{-b(1+z)\Gamma}.
\label{expo_inv}
\ee
If $W_+$ in Eq.~(\ref{gamma}) is taken to be $q=2p_T$, Eq.~(\ref{expo_inv})
can be expressed in terms of $p_T$ as
\ba
E\,{d^3\sigma\over dp^3}&=&\left[
{4N\,(1-z)^a\,\sigma_0\over\sqrt{s^2z^2-4s\,m^2}}\,
e^{b(1+z)\,m_h^2}\right]\,e^{-2b(1+z)\,m_h\,e^{-y}\,p_T}\\
\label{expo_pT_a}
&\simeq&\left[{4N\,(1-z)^a\,\sigma_0\over\sqrt{s^2z^2-4s\,m^2}}\,
e^{b(1+z)\,m_h^2}\right]\,e^{-2b(1+z)\,m_h\,p_T}\\
&\equiv&A\,e^{-B\,p_T}.\label{expo_pT}
\ea
Again $y=0$ is assumed in Eq.~(\ref{expo_pT_a}).  Average values of
$s$ and $z$ are used when comparing with experimental data so that
$A$ and $B$ are treated as constants.
It must be emphasized that $s$ is not the beam energy square but the total
energy square
of the produced mesons in the string fragmentation process.
The parameters for pion cross section are listed in Table~\ref{softpar}.
$B$ is extracted from data as shown in Figs.~\ref{pp_non}--\ref{pi0_non}.
$A$ is chosen to match the pion cross sections at the boundary of
$p_T=1\,\rm GeV$ between the soft and hard $p_T$ regions so that
\be
A=3\times10^{-28}\,e^B
\ee
in unit of $\rm cm^2/GeV^2$ for pions.  These figures suggest that the cross
sections are approximately constant in $\sqrt{s}$ and $\theta_{cm}$ in the
range of energy being parameterized $(22\,{\rm GeV}<\sqrt{s}<63\,\rm GeV)$.
Therefore the present parameterization of pion production cross sections for
$p_T<1\,\rm GeV$ is
expected to be valid for the energy range not narrower than
$20\,{\rm GeV}<\sqrt{s}<70\,\rm GeV$.
Kaons are not analyzed nor parameterized because of insuffucient data.

\section{Pion and Kaon production in the hard $p_T$ region}

The invariant cross sections of pions and kaons from proton-proton
scattering in the hard $p_T$
region $(p_T\ge1\,\rm GeV)$ are treated in this section by using
perturbative QCD.  The theoretical foundation of
the present calculation is based on the Feynman-Field parton model.
A comprehensive review of the model is given in References~\cite{field,appen}.
The cross sections are parameterized at the end.

The Feynman-Field invariant production cross section formula
incorporates the parton
distribution functions, $f_{A/a}(x_a, Q^2)$, obtained from DIS experiments
and the fragmentation functions, $D^h_q(z, Q^2)$, derived from a combination
of stochastic arguments and parameterizations of data.  The cross section
formula is given as~\cite{field}
\be
E\,{d^3\sigma\over dp^3}={1\over\pi}\sum_{a,b}\int^1_{x_a^{min}}dx_a
\int^1_{x_b^{min}}dx_b\,f_{A/a}(x_a,Q^2)\,f_{B/b}(x_b,Q^2)\,D^h_c(z_c,Q^2)\,
{1\over z_c}\,{d\sighat\over d\that},\label{csf}
\ee
with
\ba
x_a^{min}&=&{x_1\over1-x_2},\\
x_b^{min}&=&{x_a x_2\over x_a-x_1}.
\ea
Monte Carlo integration package \texttt{VEGAS} is used to calculate
Eq.~(\ref{csf}).  The parton distributions of proton is
given by the \texttt{CTEQ6} package~\cite{cteq6}.
The QCD running coupling constant, $\alpha_s(Q^2)$,
is the renormalized
coupling constant described in Reference~\cite{appen}.  A typical value of
$\Lambda=0.4\,{\rm GeV}$ is used inside $\alpha_s(Q^2)$.  The internal
scattering cross sections of the reactions $q_i\qbar_i\to gg$ and
$gg\to gg$ are excluded from the integral 
because gluons do not fragment into hadrons.
The fragmentation functions used for this calculation are the
original fragmentation functions of Feynman and Field~\cite{ff78}.
For the $pp\to\pi\,X$ reactions, the fragmentation functions are
\be
D^{\pi^0}_u(z)=D^{\pi^0}_d(z)=\left[{\beta\over2}
+\beta^2\left({1-z\over z}\right)\right]\,(n+1)\,(1-z)^n,
\ee
\be
D^{\pi^0}_s(z)=\beta^2\left({1-z\over z}\right)\,(n+1)\,(1-z)^n,
\ee
\be
D^{\pi^-}_d(z)=D^{\pi^+}_u(z)=\left[\beta
+\beta^2\left({1-z\over z}\right)\right]\,(n+1)\,(1-z)^n,
\ee
\be
D^{\pi^{\pm}}_s(z)=D^{\pi^+}_d(z)=D^{\pi^-}_u(z)=
\beta^2\left({1-z\over z}\right)\,(n+1)\,(1-z)^n,
\ee
and for $pp\to K\,X$ reactions, the fragmentation functions are
\be
D^{K^+}_u(z)=D^{K^0}_d(z)={1\over2}\,\left[\beta
+\beta^2\left({1-z\over z}\right)\right]\,(n+1)\,(1-z)^n,
\ee
\be
D^{K^+}_s(z)=D^{K^0}_s(z)=D^{K^+}_d(z)=D^{K^0}_u(z)=
{\beta^2\over2}\,\left({1-z\over z}\right)\,(n+1)\,(1-z)^n,
\ee
\be
D^{\overline{K}^0}_s(z)=D^{K^-}_s(z)=\left[\beta
+{\beta^2\over2}\,\left({1-z\over z}\right)\right]\,(n+1)\,(1-z)^n,
\ee
\be
D^{K^-}_d(z)=D^{K^-}_u(z)=D^{\overline{K}^0}_d(z)=D^{\overline{K}^0}_u(z)=
{\beta^2\over2}\,\left({1-z\over z}\right)\,(n+1)\,(1-z)^n,
\ee
where $\beta=0.4$.
The distributions of $c$, $b$ and $t$ quarks are sufficiently low that
\be
D^h_c(z)=D^h_b(z)=D^h_t(z)=0,
\ee
for any hadron $h$.  Feynman fixed $n=2$ in his original paper.  In this work,
$n$ is a parameter freely adjusted to fit data.
There is a subtlety involved in summing all the
parton contributions over $a$ and $b$ in Eq.~(\ref{csf}) that is related to
the relative probabilitistic nature of the parton distributions and our
ignorance of the number of sea quarks and gluons inside the proton.
The parton distributions are normalized to
unity so that they give only the relative distributions of the partons.
The parton distributions give only the ratios of the partons in a hadron but
not their numbers.  In order to sum over $a$ and $b$ partons in Eq.~(\ref{csf})
correctly, an integral multiplicative constant for
each of the hadrons $A$ and $B$ must be provided.
These multiplicative integral constants are not
known {\em a prior\'{i}} but are determined {\em a posterior\'{i}}
by fitting data.  In other words, Eq.~(\ref{csf}) can be modified as
\ba
E\,{d^3\sigma\over dp^3}&=&{N_A\,N_B\over\pi}\,
\sum_{\{a,b\}}\,\int^1_{x_a^{min}}dx_a
\int^1_{x_b^{min}}dx_b\nonumber\\
&&\quad\times f_{A/a}(x_a,Q^2)\,f_{B/b}(x_b,Q^2)\,
D^h_c(z_c,Q^2)\,{1\over z_c}\,{d\sighat\over d\that},\label{csf2}
\ea
where $N_A$ and $N_B$ are the multiplicative constants corresponding
to $f_{A/a}(x_a, Q^2)$
and $f_{B/b}(x_b, Q^2)$ respectively and $\sum_{\{a,b\}}$ is the sum over
the parton {\em types} instead of a sum over the partons {\em per se}.
If $A=B$, the overall multiplicative constant, $N_A\,N_B$, is an integer
square.  If $A\ne B$, the overall constant, $N_A\,N_B$,
is still an integer.  In the case of fitting pQCD calculations
to experimental data of a $pp\to\pi X$ reaction, a factor of 100 is missing
if one simply sums over the parton types.  It implies that the multiplicative
constant for the parton distributions of proton is $N_p=10$.  The cross
sections for $K$ production is approximately half of that of $\pi$
production.  It implies that the multiplicative constant for the
$pp\to K\,X$ reaction may be $N_p=7$.  For the purpose parameterizing the shape
of the kaon production cross section, an exact scale is not required.
Therefore
$N_p$ in the $pp\to KX$ reactions is arbitrarily set to be the same as that of
$pp\to\pi X$ reactions at $N_p=10$.  This choice is adequate because
fits to experimental data of kaons are not being pursued in this work due to
the lack of experimental data for kaons.

It is observed that the invariant production cross sections
have the same basic shape regardless of the reactions, {\em i.e.} an
exponentially decaying function of the form $\exp(-\alpha x^\beta)$ at low
$p_T$ and a suppression at high $p_T$ which drops off to zero before the edge
of suppresion at $p_T\le\sqrt{s}/2$.  The cross
section is at its maximum at $p_T=0$ and decreases monotonically in $p_T$.
The Feynman-Field code used in this calculation assumes that
$Q^2=4p_T^2$.  In other words, by combining the previous two statements, the
cross section is at its maximum at $Q^2=0$ and decreases monotonically in
$Q^2$.  This observation indicates that hadron fragmentation
is more favorable at low $Q^2$ in that a parton preserves more kinetic
energy to be made available for hadron fragmentation.  It is also observed
that the cross section is suppressed at high $p_T$ and that the edge of
suppression of the cross section is $\sqrt{s}/2$ at low
$p_T$ and gradually increases toward high $p_T$.  The reason for this
phenomenon is mostly due to the choice of $Q^2=4p_T$ such that
$s\ge Q^2$ or equivalently $\sqrt{s}/2\ge p_T$.  It turns out that the
edge of suppression along $p_T$ is a function of $\sqrt{s}$.  The comments
made so far apply to any angle $\theta_{cm}$ when $\sqrt{s}$ is replaced by
$\sin\theta\sqrt{s}$.  The basic features of the graphs of the invariant
cross section is discussed in Reference~\cite{appen}.

\begin{figure}
\includegraphics[scale=0.6]{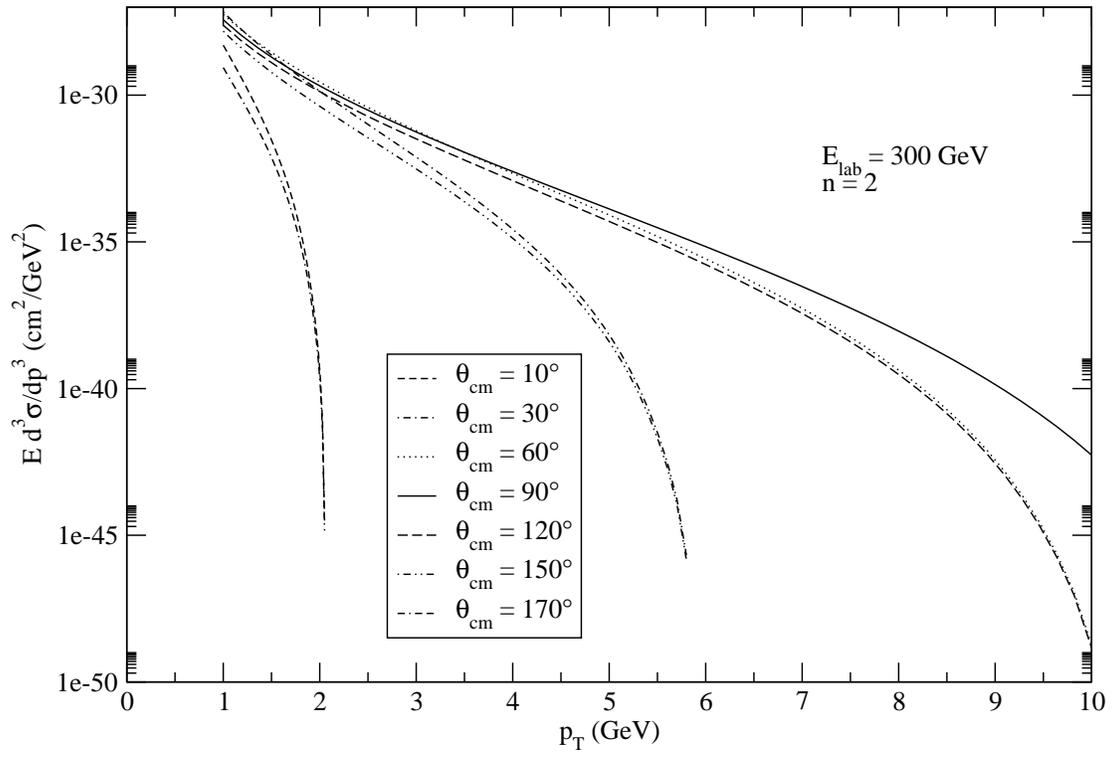}
\caption{\label{test_deg}
A sample plot of the Feynman-Field model for $pp\to\pi^+X$
at different angles.  The plot shows that the
invariant cross section is symmetric around $\theta_{cm}=90^{\circ}$
and is suppressed at $\theta_{cm}\ne90^{\circ}$.}
\end{figure}

A lot of effort has been invested in parameterizing
$E\,d^3\sigma/dp^3$ from experimental data for use in the \texttt{HZETRN}
code~\cite{steve}.  This work takes a different approach by parameterizing
theory.  Experimental data are used merely as a means to fine-tune the
parameterizations.  
Monte Carlo integration is the fastest numerical integration scheme available
but it is not faster than an analytic formula.
On the other hand, the explicit computation of the double integrals in
the Feynman-Field model is a daunting task if tractable at all.
These constraints motivate the present parameterization
scheme.  The method of finding the parameters is mostly one of trial-and-error.
Guesses of the appropriate functions are made for different parts of the curve.
The pieces are put back together at the end and refit until the
parameterization has the desired global properties.  The final form of
paramterized cross section formula is found to be
\ba
E\,{d^3\sigma\over dp^3}(\sqrt{s},\,p_T,\,\theta_{cm})&=&A\,
e^{-35.4\beta\,(p_T^\beta-1)}\nonumber\\
&&\times\left[\exp\left(1-\csc\left(\left(
{p_T-1\over a\left({\sin\theta_{cm}\sqrt{s}\over2}-1\right)}\right)^{b\beta}
{\pi\over2}\right)\right)\right]^{1\over b\beta},\label{para}
\ea
for $p_T-1<a\left({\sin\theta_{cm}\sqrt{s}\over2}-1\right)$.  It is taken that
\be
E\,{d^3\sigma\over dp^3}=0,
\ee
for $p_T-1\ge a\left({\sin\theta_{cm}\sqrt{s}\over2}-1\right)$.
$A$ is a scale factor such that
\be
A=E\,{d^3\sigma\over dp^3}(\sqrt{s},\,p_T=1,\,\theta_{cm}=90^{\circ}).
\ee
The functions $a$ and $\beta$ are
\be
a=2-e^{-\alpha\,\sin\theta_{cm}\sqrt{s}},
\ee
\be
\beta=\beta_0+\beta_1\,\left(\sin\theta_{cm}\sqrt{s}\right)^{-\beta_2},
\ee
and the parameters $a$, $b$, $\beta_0$, $\beta_1$ and $\beta_2$ are freely
adjusted
to fit the curve.  The unit of energy-momentum is GeV for the present
parameterization.  $A$ is determined by varying $n$ to fit the data with
the Feynman-Field calculation.
The first exponential in
Eq.~(\ref{para}) controls the shape of the curve at low $p_T$ and
the square bracket term controls the suppression at high $p_T$.  The function
$a$ shifts the edge of the suppression so that the edge is located at
$\sqrt{s}/2$ at low $p_T$ and $\sqrt{s}$ at high $p_T$.  There is a
threshold $p_T\sim0.2\,{\rm GeV}$ set by the \texttt{CTEQ6} package.  
In addition, non-perturbative effects become more prominent in the soft $p_T$
region $(p_T<1\,\rm GeV)$ so that the pQCD code cannot be applied there.
For these reasons, the present parameterization focuses on the region
$p_T\ge1\,{\rm GeV}$.  The rapidity
distributions of hadron production is typically symmetric around $y=0$.
From~\cite{wong94}
\be
\eta=-\ln[\tan(\theta/2)]=
{1\over2}\,\ln\left[{\sqrt{m^2_T\cosh^2y-m^2}+m_T\sinh y\over
\sqrt{m^2_T\cosh^2y-m^2}-m_T\sinh y}\right]
\ee
and
\be
y={1\over2}\,\ln\left({E+p_z\over E-p_z}\right),
\ee
it can be easily shown that $y=0$, $x_L=p_z/(p_z)_{max}=0$ and
$\theta_{cm}=90^{\circ}$ are the same statements.  Many experiments
average the data over a range of $y$ or $x_L$ symmetric around zero.  In
these cases, the average center-of-momentum angle,
$\overline{\theta}_{cm}$, would be $90^{\circ}$, which happens to be the
most prominent contribution according to Fig.~\ref{test_deg}.
Hadron productions in
space radiation problems are mostly in the forward direction or equivalently
$\theta=0$.  In highly relativistic regimes,
$\overline{\theta}_{cm}=90^{\circ}$ is transformed to a small
$\theta_{lab}$.  For example, $\theta_{cm}=90^{\circ}$ is
equivalent to $\theta_{lab}=9.17^{\circ}$ at
$\sqrt{s}=70\,{\rm GeV}$~\cite{abramov80}.  The scattering angle $\theta_{lab}$
decreases as
$\sqrt{s}$ increases.  At sufficiently high $\sqrt{s}$, $\theta_{lab}$
is effectively zero so that a one-dimensional transport code is
justified and that the largest angular contribution of the cross section is
that at $\theta_{cm}=90^{\circ}$.

\begin{figure}
\includegraphics[scale=0.6]{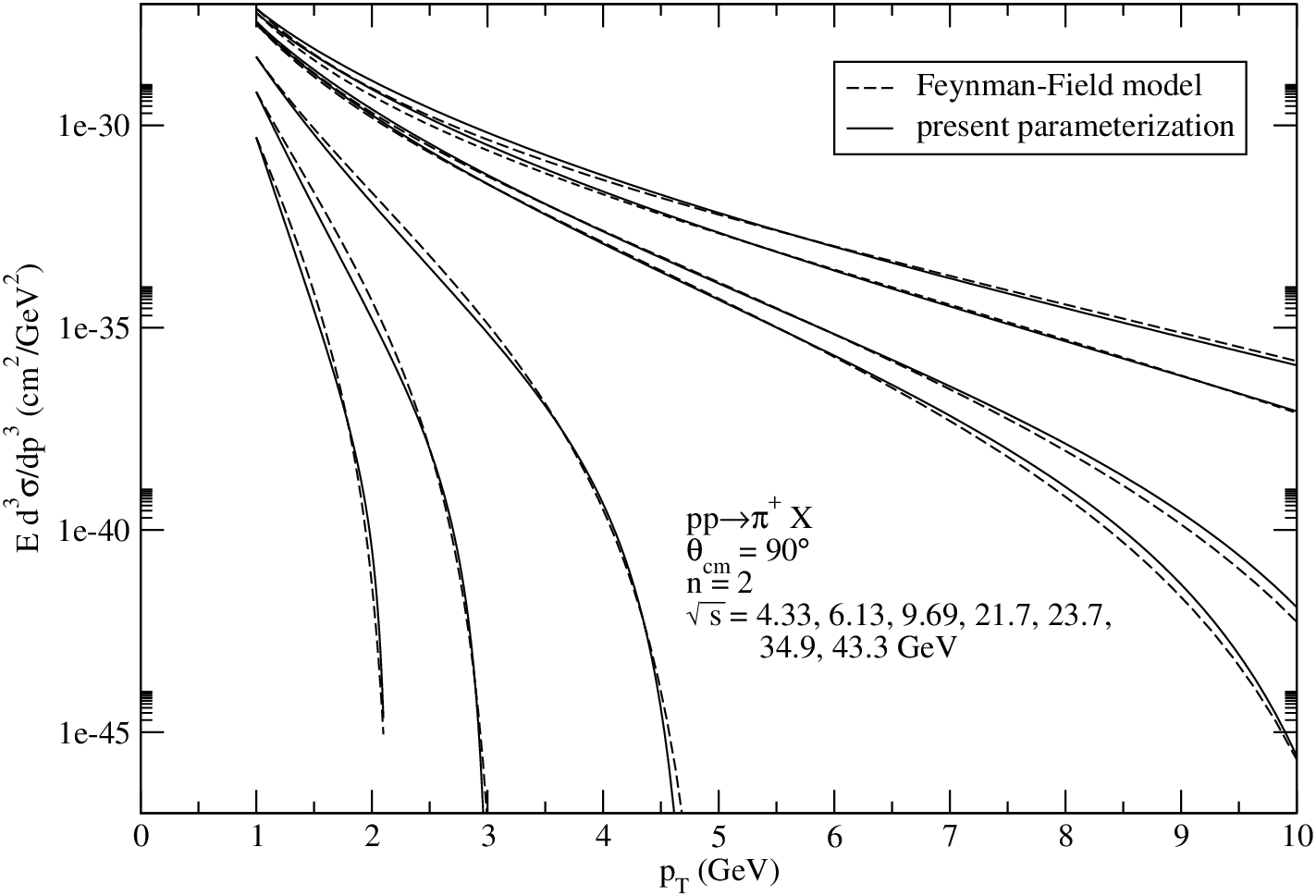}
\caption{\label{pip_en}
Parameterization of the Feynman-Field model for $pp\to\pi^+X$
at $\theta_{cm}=90^{\circ}$ and $n=2$.  The values of $\sqrt{s}$ listed
in the figure match the curves from left to right.}
\end{figure}

\begin{figure}
\includegraphics[scale=0.6]{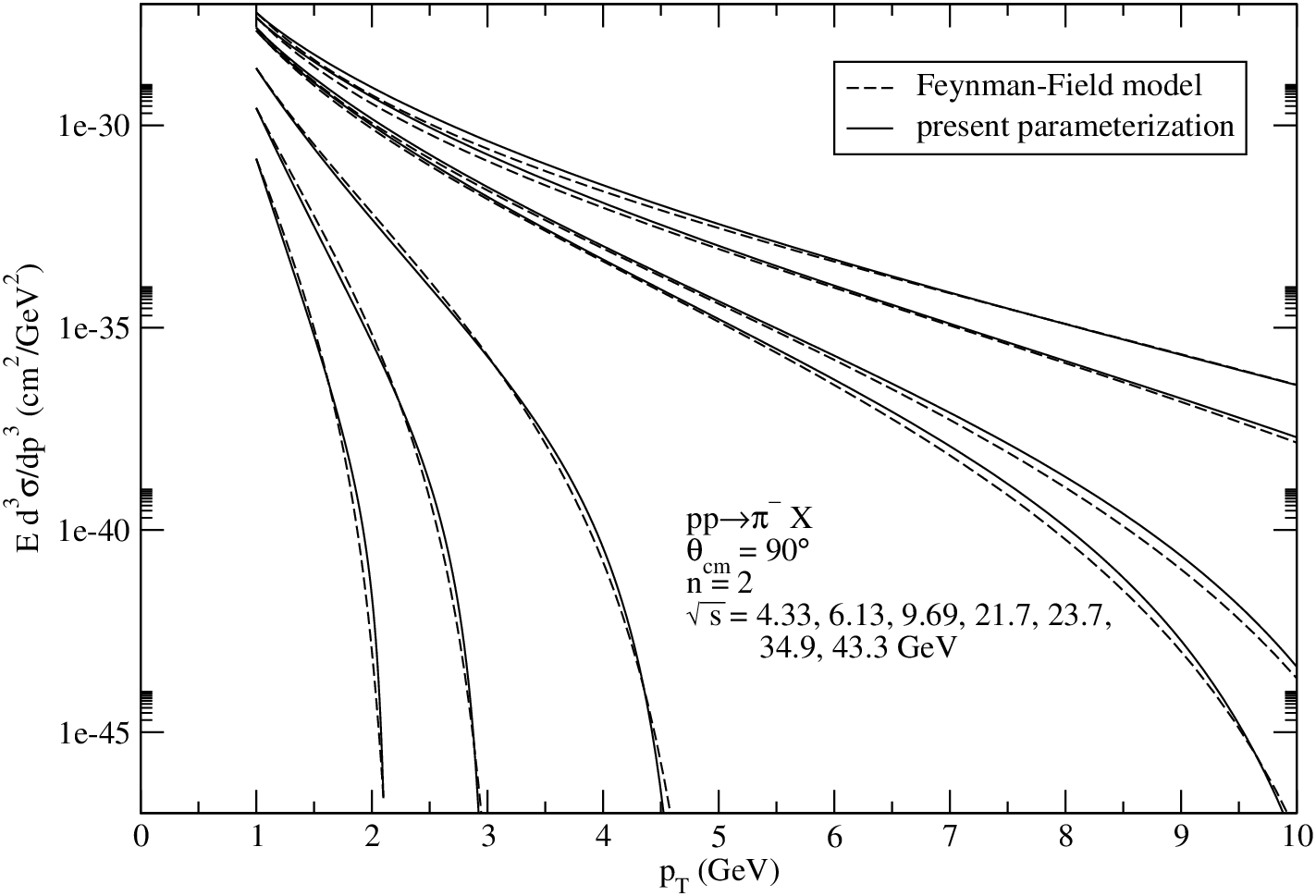}
\caption{\label{pim_en}
Parameterization of the Feynman-Field model for $pp\to\pi^-X$
at $\theta_{cm}=90^{\circ}$ and $n=2$.  The values of $\sqrt{s}$ listed
in the figure match the curves from left to right.}
\end{figure}

\begin{figure}
\includegraphics[scale=0.6]{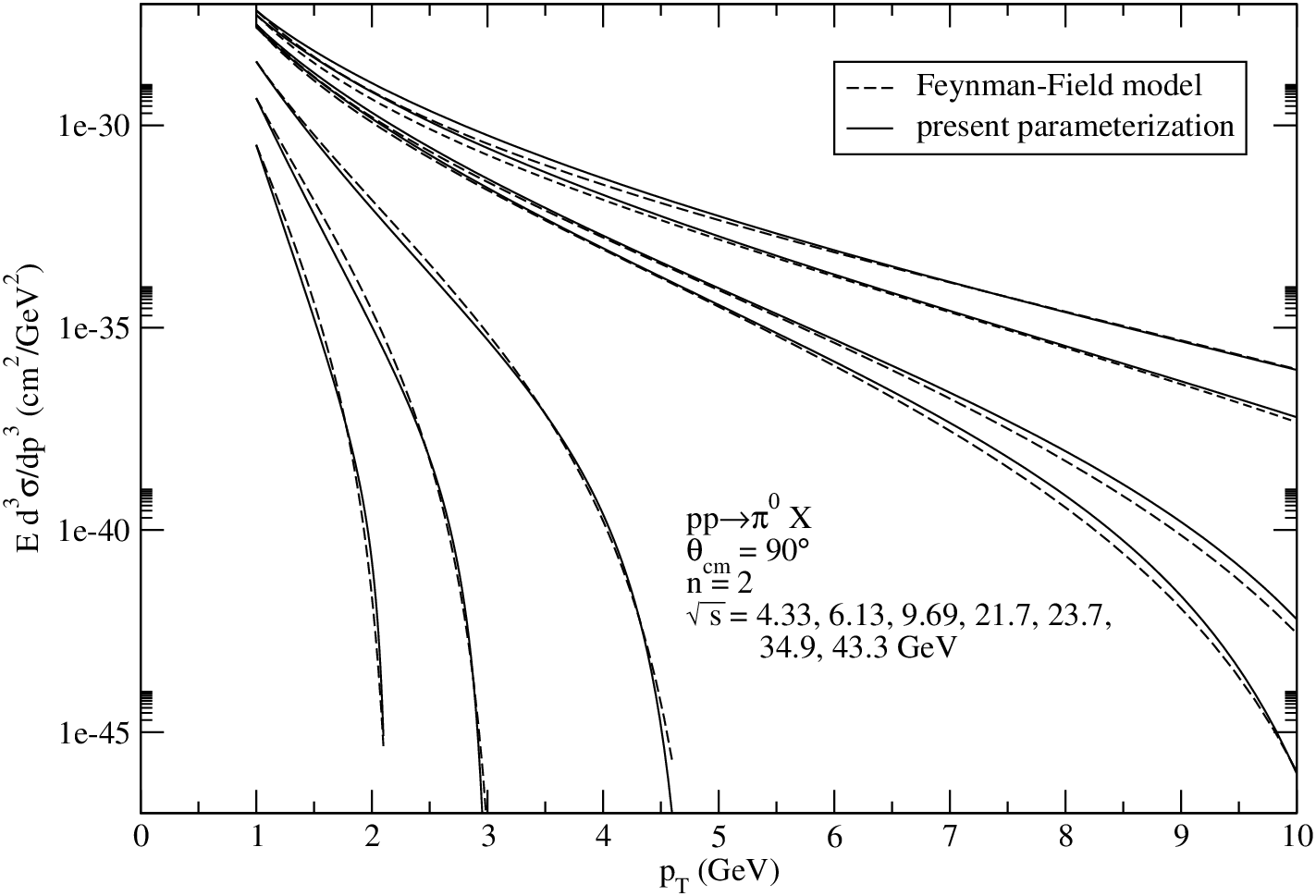}
\caption{\label{pi0_en}
Parameterization of the Feynman-Field model for $pp\to\pi^0X$
at $\theta_{cm}=90^{\circ}$ and $n=2$.  The values of $\sqrt{s}$ listed
in the figure match the curves from left to right.}
\end{figure}

\begin{figure}
\includegraphics[scale=0.6]{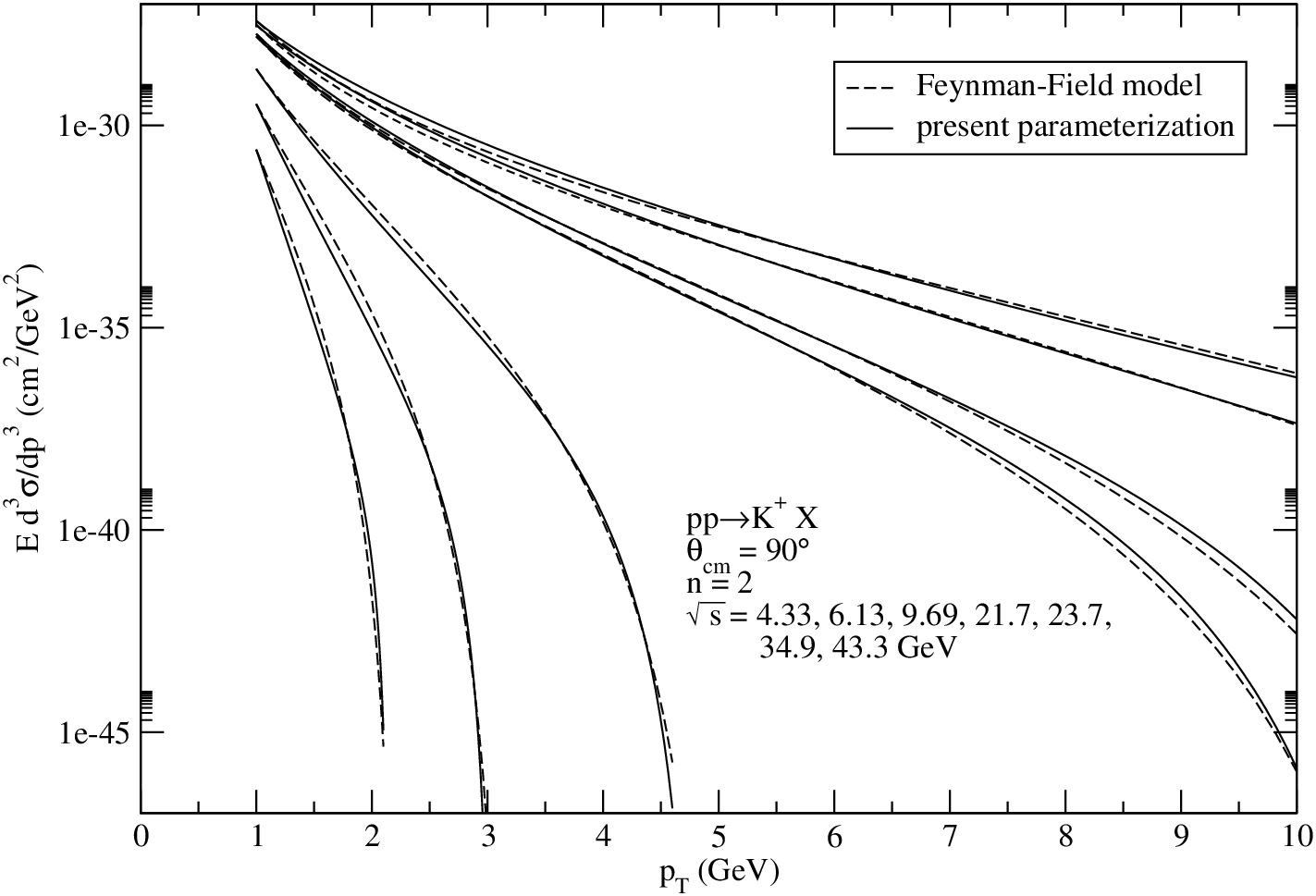}
\caption{\label{kp_en}
Parameterization of the Feynman-Field model for $pp\to K^+X$
at $\theta_{cm}=90^{\circ}$ and $n=2$.  The values of $\sqrt{s}$ listed
in the figure match the curves from left to right.}
\end{figure}

\begin{figure}
\includegraphics[scale=0.6]{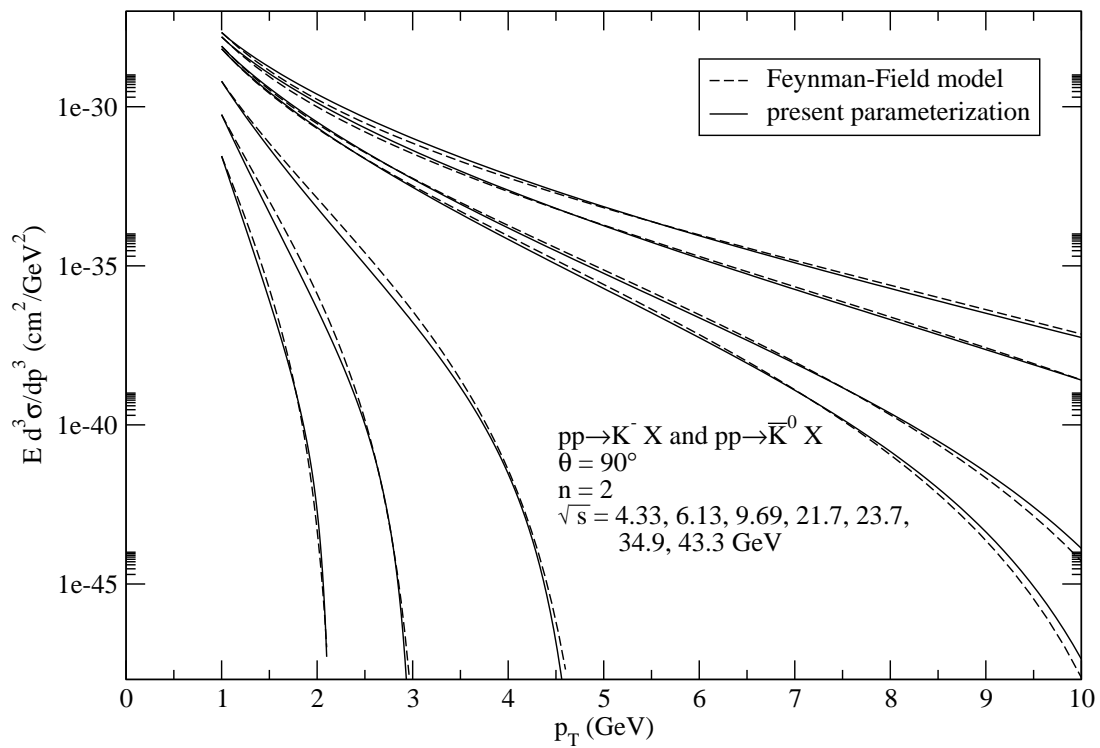}
\caption{\label{km_en}
Parameterization of the Feynman-Field model for the reactions
$pp\to K^-X$ and $pp\to \overline{K}^0X$
at $\theta_{cm}=90^{\circ}$ and $n=2$.    The values of $\sqrt{s}$ listed
in the figure match the curves from left to right.
$K^-$ and $\overline{K}^0$
have the same fragmentation functions and hence the same
Feynman-Field cross section.  However their experimental cross
sections may not be the same.}
\end{figure}

\begin{figure}
\includegraphics[scale=0.6]{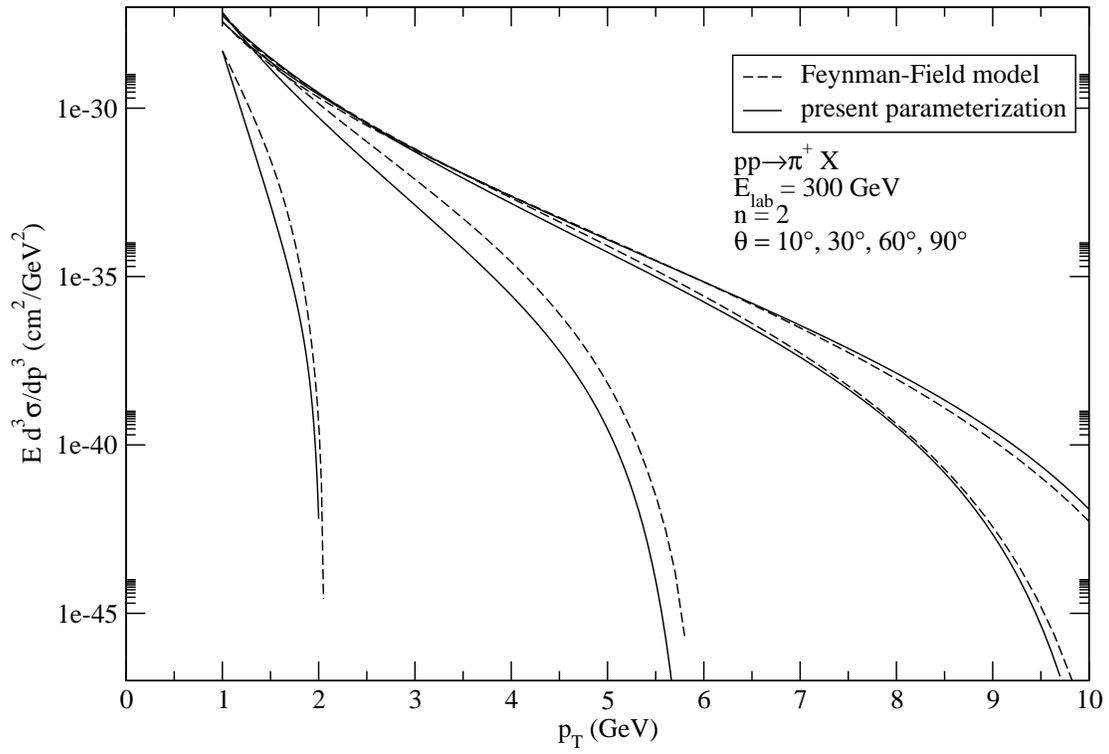}
\caption{\label{pip_deg_c}
Parameterization of the Feynman-Field model for $pp\to\pi^+X$
at various angles and $n=2$.  The values of the angles listed in the
figures match the curves from left to right.}
\end{figure}

\begin{figure}
\includegraphics[scale=0.6]{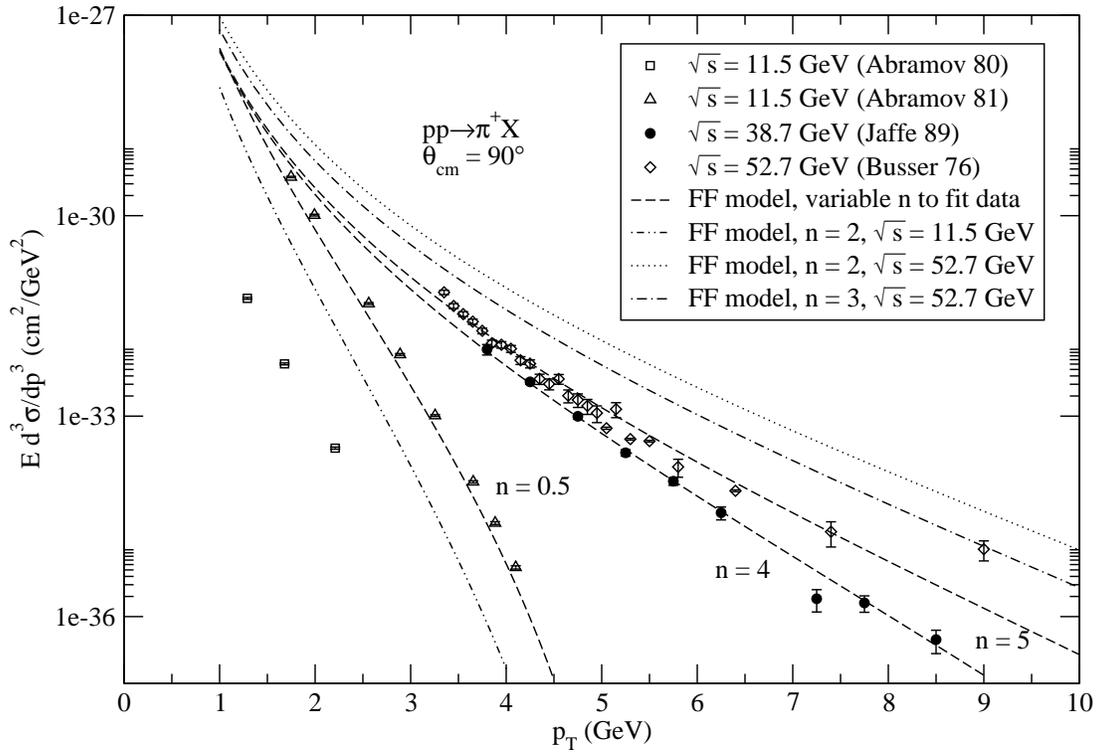}
\caption{\label{pip_demo}
Comparisons of the Feynman-Field model fit for $pp\to\pi^+X$ for various
$n$ at $\theta_{cm}=90^{\circ}$.  The references of the data sets Abromov80,
Abramov81, Jaffe89 and
Busser76 are~\cite{abramov80}, \cite{abramov81}, \cite{jaffe89} and
\cite{busser76} respectively.}
\end{figure}

\begin{figure}
\includegraphics[scale=0.6]{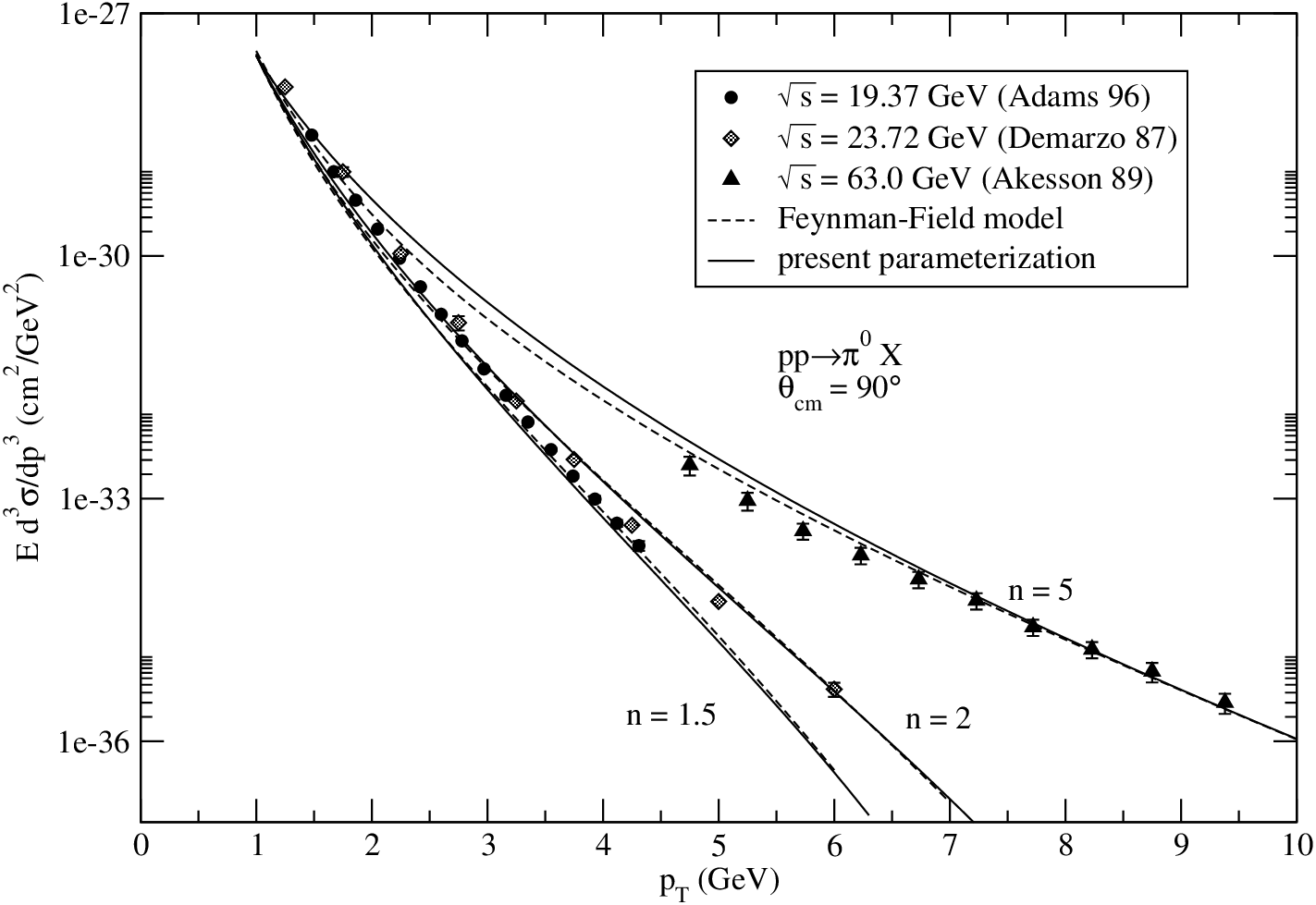}
\caption{\label{pi0}
Comparisons of pQCD and parameterized cross sections with
$pp\to\pi^0X$ experimental data at $\overline{\theta}_{cm}=90^{\circ}$.
The parameter $n$ in fragmentation functions of the Feynman-Field model
are adjusted freely as shown in the graph to fit
the data.  The references of the data sets Adams96, Demarzo87 and
Akesson89 are~\cite{adams96}, \cite{demarzo87} and \cite{akesson89}
respectively.}
\end{figure}

\begin{figure}
\includegraphics[scale=0.6]{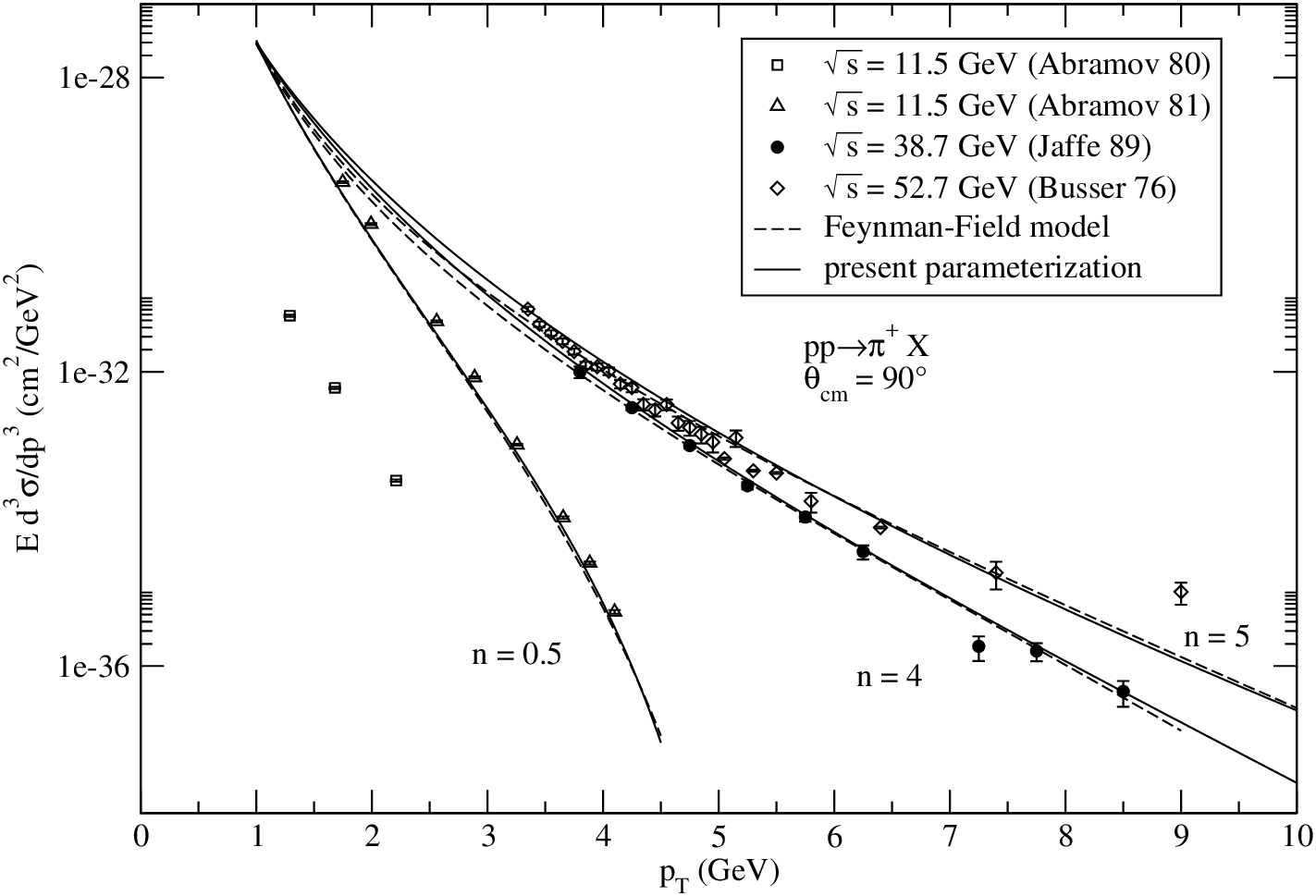}
\caption{\label{pip}
Comparisons of pQCD and parameterized cross sections with
$pp\to\pi^+X$ experimental data at $\overline{\theta}_{cm}=90^{\circ}$.
The parameter $n$ in fragmentation functions of the Feynman-Field model
are adjusted freely as shown in the graph to fit
the data.  The references of the data sets Abromov80, Abramov81, Jaffe89 and
Busser76 are~\cite{abramov80}, \cite{abramov81}, \cite{jaffe89} and
\cite{busser76} respectively.}
\end{figure}

\begin{figure}
\includegraphics[scale=0.6]{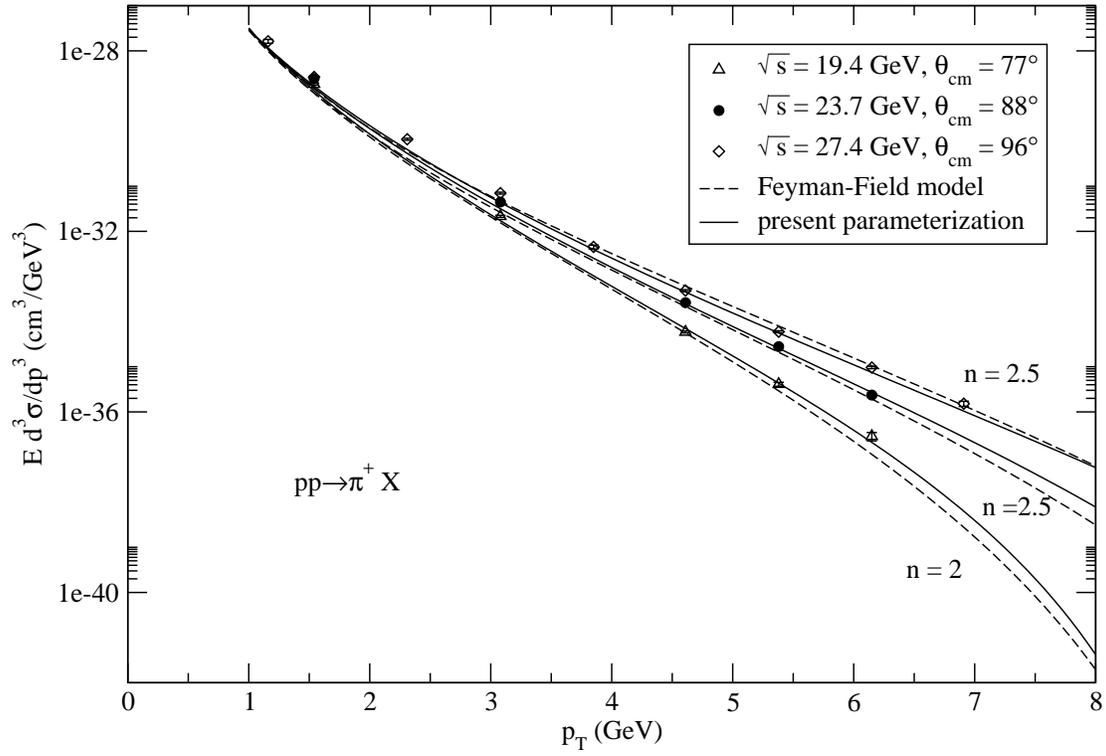}
\caption{\label{pip_deg}
Comparisons pQCD and parameterized cross sections with
$pp\to\pi^+X$ experimental data at various energies and angles.
The parameter $n$ in fragmentation functions of the Feynman-Field model
are adjusted freely as shown in the graph to fit
the data.  The experimental data set is published in
Reference~\cite{antreasyan77}.}
\end{figure}

Figs.~\ref{pip_en}-\ref{km_en} show the goodness of fit between the
parameterized cross sections and the pQCD calculations at
$\theta_{cm}=90^{\circ}$ and Fig.~\ref{pip_deg_c} shows the goodness of
fit at various angles.  Apparently the same parameterization
of $\beta$ works for all types of pions and kaons
(see Figs.~\ref{pip_en}--\ref{km_en}).
The Feynman-Field model fits data
well when $n$ is allowed to vary as it is illustrated in Fig.~\ref{pip_demo}.
Generally speaking, the parameterization of the model at a fixed value of
$n=2$ does not always fit data well.  For this reason, data are
parameterized separately by reparameterizing $\beta$.  The parameterizations
of pions are illustrated in Figs.~\ref{pi0}--\ref{pim}.  Data of kaons
are fragmented so that only the theory (but not the data) is parameterized.
The parameters are tabulated in Table~\ref{partable}.  The parameterization
of data is distinguished from that of theory by labelling the
paramters of the former as $\beta$ and those of the latter $\beta'$.  All other
parameters are the same for both data and theory.
A sample graph of the fit between
the present parameterization and data with angular dependence is
shown in Fig.~\ref{pip_deg}.

The parameters in Table~\ref{partable} are obtained by fitting curves
by hand.  Curve-fitting algorithm such as the
Levenberg-Marquardt does not work well because of the presence of
singular matrices inherent in the present model.  At this point,
there is not a cleverer method to find these parameters automatically.  When
the parameters are obtained by hand, several data sets of
$E\,d^3\sigma/dp^3$ versus $p_T$ at different energies with $p_T>1\,\rm GeV$
and
$\overline{\theta}_{cm}=90^{\circ}$ are needed.  In general, more data sets
means smaller $\chi^2$.  In the case of parameterizing theory, the Monte
Carlo program can generate as many theoretical data sets as needed.  In the
case of parameterizing experiment, data are generally fragmented except
those of pions.
At least 3 experimental data sets are needed to fit $\beta$.  Pion data
are generally quite copious.  In the case of kaons, the scarcity
of data prevents the parameterization of their experimental fits.  The present
parameterization of pion production cross sections for $p_T>1\,\rm GeV$ is
expected to be valid over $10\,{\rm GeV}<\sin\theta_{cm}\sqrt{s}<70\,\rm GeV$,
which is given by the range of energy of the experimental data used in this
analysis. 

Most experiments agree with the shape of theoretical curves.  However there are
diagreements in the magnitudes of the cross sections among experimental data
sets, sometimes even by the same authors.  Figs.~\ref{pip} and \ref{pim}
show that Abramov {\em et al.} published 2 sets of pion data in identical
energy regimes in 2 consecutive years that are different by 3 orders of
magnitude\cite{abramov80,abramov81}.  This phenomenon occurs
quite regularly in $E\,d^3\sigma/dp^3$ data.  Some discretion must be
exercised in choosing the experimental data to parameterize.

\section{Conclusion}
\begin{figure}
\includegraphics[scale=0.6]{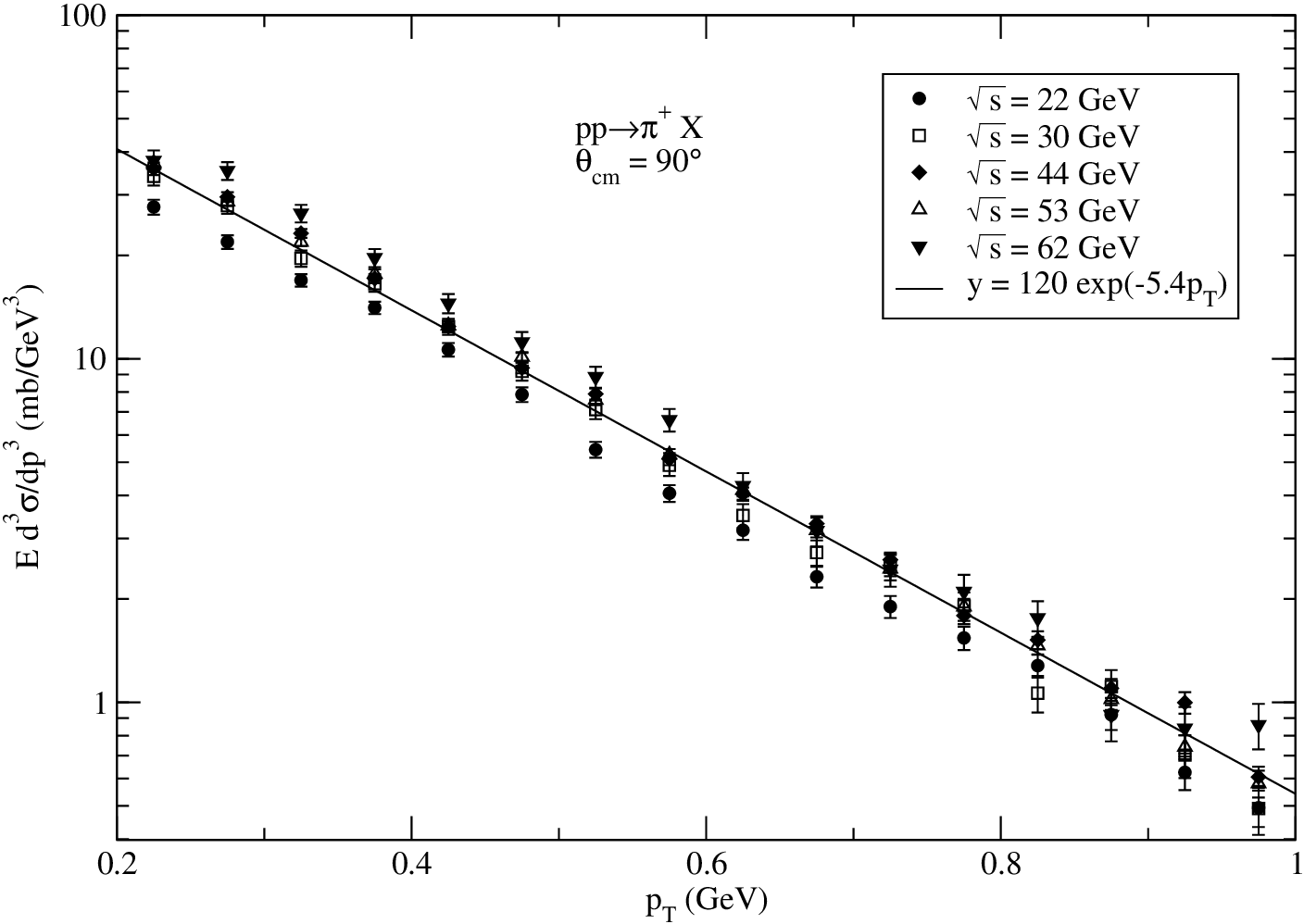}
\caption{\label{pp_non}
Comparison between experiment and theory of $E\,d^3\sigma/dp^3$ 
for $pp\to\pi^+\,X$ in the
soft $p_T$ region $(p_T<1 \rm GeV)$.
The data are published in Reference~\cite{banner77}.}
\end{figure}

\begin{figure}
\includegraphics[scale=0.6]{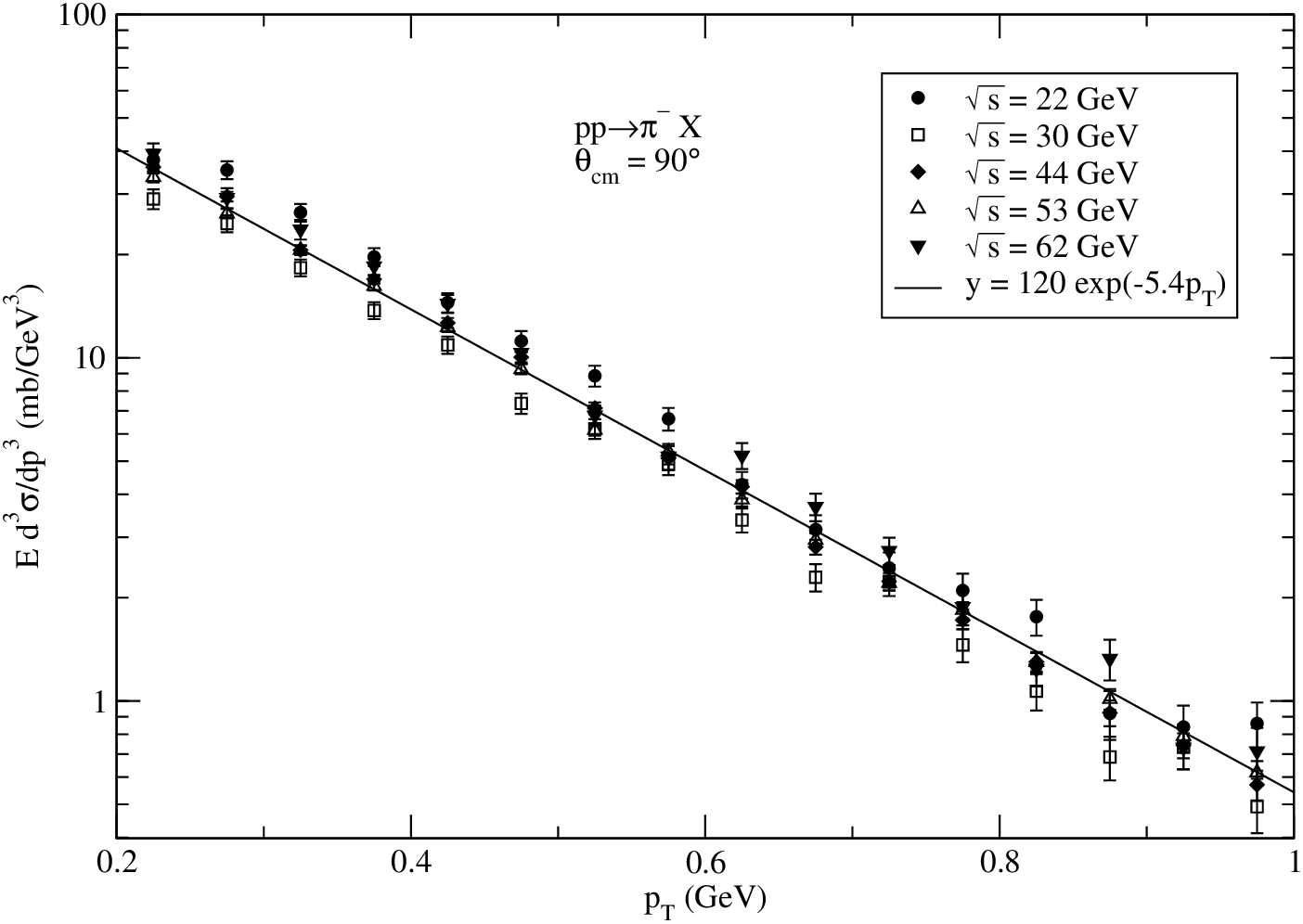}
\caption{\label{pm_non}
Comparison between experiment and theory of $E\,d^3\sigma/dp^3$
for $pp\to\pi^-\,X$ in the
soft $p_T$ region $(p_T<1 \rm GeV)$.
The data are published in Reference~\cite{banner77}.}
\end{figure}

\begin{figure}
\includegraphics[scale=0.6]{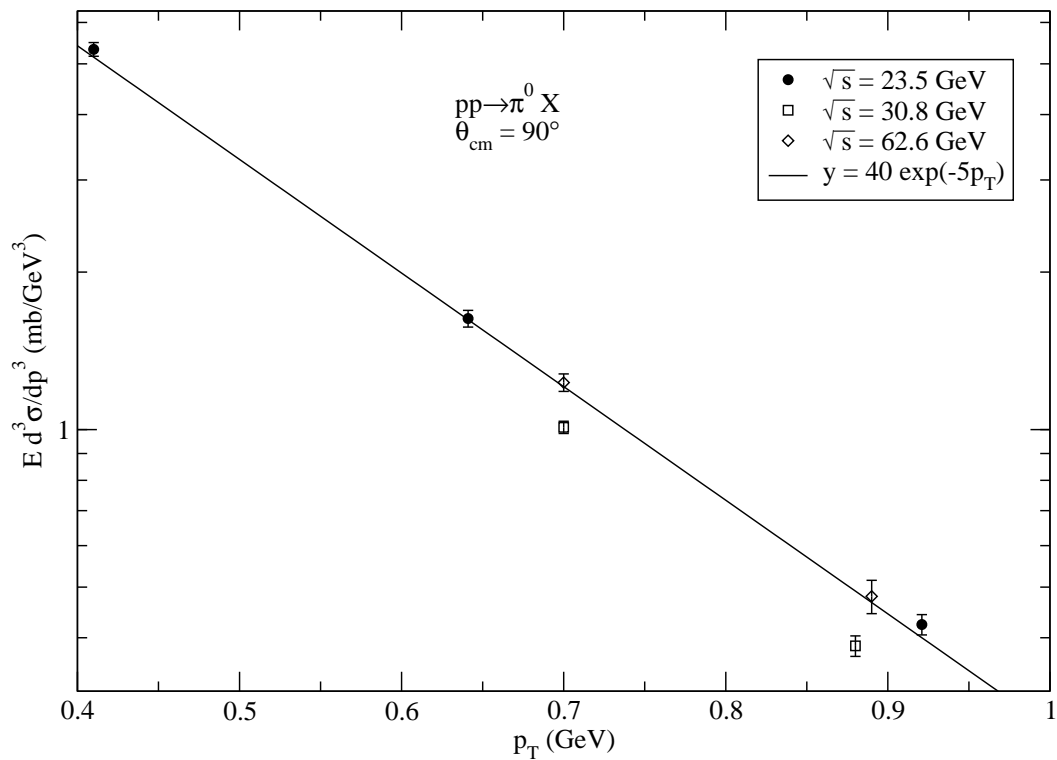}
\caption{\label{pi0_non}
Comparison between experiment and theory of $E\,d^3\sigma/dp^3$
for $pp\to\pi^0\,X$ in the
soft $p_T$ region $(p_T<1\,\rm GeV)$.
The data are published in Reference~\cite{eggert75}.}
\end{figure}

This work shows how to use the Lund model to calculate the invariant
production cross section non-perturbatively in the soft $p_T$ region.
This model predicts that the
functional form of the cross section in this sector is a simple exponential.
This prediction is confirmed by experiment in the case of pions
for $\theta=90^\circ$.  The cross section formula in Eq.~(\ref{expo_pT})
has no angular nor energy dependence.  Although the prediction of angular
independence is not yet confirmed by data, energy independence is shown
to be approximately correct according to the Figs.~\ref{pp_non}-\ref{pi0_non}.
The Feynman-Field model is a work horse
for calculating inclusive cross sections and is generally accepted to be
an accurate description of data in the high $p_T$ region.  Several
typical assumptions of pQCD have been adopted in the present calculation
that are not necessarily unique, such as
the forms of $Q^2$ and the fragmentation functions.  The
form of momentum transfer is assumed to be $Q^2=4p_T^2$ in the present work.
In principle, other guesses, such as
\be
Q^2={2\shat\that\uhat\over\shat^2+\that^2+\uhat^2},\label{other}
\ee
can also used.  Nevertheless it is unlikely that the use of Eq.~(\ref{other})
in the Feynman-Field model
will drastically change its predictions.  More experimental data of kaons
are needed to determine the corresponding paramterizations.  Otherwise, the
parameterization procedure can in principle be applied to all types of mesons. 
The production cross sections of baryons are also needed for a realistic
nuclear code.  It is not yet clear that the Lund model and Feynman-Field can
be easily modified to incorporate baryon production.  Eventually
production cross sections of hadrons from $pA$ and $AA$ collisions must also be
included in the transport code.
The phenomena of heavy nuclei collisions are much more complicated because
many-body effects such as the EMC effect, Cronin effect, nuclear shadowing,
jet quenching and gluon-plasma phase transition have to be considered.
It is hoped that the present work provide some basic ideas for more
complicated calculations to be undertaken in the future.

\begin{acknowledgments}
This work was supported in part by NASA grant NCC-1-354.
\end{acknowledgments}

\begin{figure}
\includegraphics[scale=0.6]{k0_en.eps}
\caption{\label{k0_en}
Parameterization of the Feynman-Field model for $pp\to K^0X$
at $\theta_{cm}=90^{\circ}$ and $n=2$.  The values of $\sqrt{s}$ listed
in the figure match the curves from left to right.}
\end{figure}

\begin{figure}
\includegraphics[scale=0.6]{pim.eps}
\caption{\label{pim}
Comparisons of pQCD and parameterized cross sections with
$pp\to\pi^-X$ experimental data at $\overline{\theta}_{cm}=90^{\circ}$.
The parameter $n$ in fragmentation functions of the Feynman-Field model
are adjusted freely as shown in the graph to fit
the data.  The references of the data sets Abromov80, Abramov81, Jaffe89 and
Busser76 are~\cite{abramov80}, \cite{abramov81}, \cite{jaffe89} and
\cite{busser76} respectively.}
\end{figure}

\begin{table}
\caption{Parameters of the parameterized cross section formula in the
soft $p_T$ region $(p_T<1\,\rm GeV)$.  $A$ is
chosen to match the pion cross sections at the boundary between the soft and
hard $p_T$ regions and has the unit $\rm cm^2/GeV^2$.  $B$ is extracted
from data as shown in Figs~\ref{pp_non}--\ref{pi0_non}.}
\begin{ruledtabular}
\begin{tabular}{crrr}
& $\pi^0$ & $\pi^+$ & $\pi^-$ \\
\hline
$A$ & 4.45E-26    & 6.64E-26    & 6.64E-26    \\
$B$ & 5.0  & 5.4  & 5.4 \\
\end{tabular}
\end{ruledtabular}
\label{softpar}
\end{table}

\begin{table}
\caption{Parameters of the parameterized cross section formula.  $A$ is
chosen to fit data and has the unit $\rm cm^2/GeV^2$.  $\beta_i$ parameterizes
data while $\beta'_i$ parameterizes the Feynman-Field model at $n=2$.
$\beta_i$ of kaons are not parameterized because of insufficient data.}
\begin{ruledtabular}
\begin{tabular}{crrrrrrr}
& $\pi^0$ & $\pi^+$ & $\pi^-$ & $K^0$ & $K^+$ & $K^-$ & $\overline{K}^0$ \\
\hline
$\beta'_0$ & 0.33    & 0.33    & 0.33    & 0.33   & 0.33   & 0.33  & 0.33 \\
$\beta'_1$ & 1.9339  & 1.9339  & 1.9339  & 1.9339 & 1.9339 & 1.7931 & 1.7931\\
$\beta'_2$ & 1.0558  & 1.0558  & 1.0558  & 1.0558 & 1.0558 & 0.9849 & 0.9849 \\
$\alpha$   & 4.855E-3  & 4.855E-3 & 4.855E-3
& 4.855E-3 & 4.855E-3 & 7.5E-3 & 7.5E-3\\
$b$        & 0.98    & 1.00    & 0.90    & 1.00   & 0.90   & 0.85  & 0.85 \\
$A$        & 3e-28   & 3e-28   & 3e-28   & -      & -      & -    &-  \\
$\beta_0$  & 0.30    & 0.20    & 0.20    & -      & -      & -    &-  \\
$\beta_1$  & 0.3337  & 0.3228  & 0.3510  & -      & -      & -    &-  \\
$\beta_2$  & 0.3774  & 0.1472  & 0.1815  & -      & -      & -    &-  \\
\end{tabular}
\end{ruledtabular}
\label{partable}
\end{table}

\end{document}